\documentclass[journal,twocolumn]{IEEEtran}

\usepackage{palatino}
\usepackage{mathtools}
\usepackage{url}
\usepackage{amsmath}
\usepackage{amssymb}
\usepackage{wrapfig}
\usepackage{framed}
\usepackage{mdframed}
\usepackage{verbatim}
 \usepackage{amsthm}
 \usepackage{cite}
\usepackage{subfigure}
\usepackage{palatino}
\usepackage{mathtools}
\usepackage{url}
\usepackage{amsmath}
\usepackage{amssymb}
\usepackage{wrapfig}
\usepackage{framed}
\usepackage{mdframed}
\usepackage{bbm}
\usepackage{verbatim}\usepackage{subfigure}
\usepackage[utf8]{inputenc}
\usepackage{amssymb}
\usepackage{amsthm}
\usepackage{dsfont}
\usepackage[normalem]{ulem}

\usepackage{bbm}
\usepackage{subfigure}
\usepackage{times}
\usepackage{multirow}
\usepackage[final]{graphicx}
\usepackage{upgreek}
\usepackage{latexsym,amssymb,mathrsfs}
\usepackage{comment}
\usepackage{url}

\usepackage{algorithmicx}
\usepackage[ruled]{algorithm}
\usepackage{algpseudocode}
\usepackage{algpascal}
\usepackage{algc}
\usepackage{xcolor}
\usepackage{cite}

 \usepackage{cite}

\newtheorem{defi}{Definition}

\title{A Survey of Local Differential Privacy and Its Variants}
\author{\IEEEauthorblockN{Likun~Qin, Nan~ Wang, Tianshuo Qiu \\
}
\IEEEauthorblockA{Department of Electrical and Computer Engineering}\\
\IEEEauthorblockA{Shandong University, Jinan, China\\
}
}
\begin{document}
\maketitle

\begin{abstract}
The introduction and advancements in Local Differential Privacy (LDP) variants have become a cornerstone in addressing the privacy concerns associated with the vast data produced by smart devices, which forms the foundation for data-driven decision-making in crowdsensing. While harnessing the power of these immense data sets can offer valuable insights, it simultaneously poses significant privacy risks for the users involved. LDP, a distinguished privacy model with a decentralized architecture, stands out for its capability to offer robust privacy assurances for individual users during data collection and analysis. The essence of LDP is its method of locally perturbing each user's data on the client-side before transmission to the server-side, safeguarding against potential privacy breaches at both ends. This article offers an in-depth exploration of LDP, emphasizing its models, its myriad variants, and the foundational structure of LDP algorithms.
\end{abstract}

\section{Introduction}

Collecting and analyzing data introduces significant privacy concerns because it often includes sensitive user information. With the advent of sophisticated data fusion and analysis methods, user data becomes even more susceptible to breaches and exposure in this era of big data. For instance, by studying appliance usage, adversaries can deduce daily routines or behaviors of individuals, like when they are home or their specific activities such as watching TV or cooking. It's crucial to prioritize the protection of personal data when gathering information from diverse devices. Currently, the European Union (EU) has released the GDPR\cite{GDPR2016a}, which oversees EU data protection laws for its citizens and outlines the specifics related to the handling of personal data. Similarly, the U.S. National Institute of Standards and Technology (NIST) is in the process of crafting privacy frameworks. These frameworks aim to more effectively recognize, evaluate, and address privacy risks, enabling individuals to embrace innovative technologies with increased trust and confidence\cite{Facebook,free_lunch}.

From a privacy-protection standpoint, differential privacy (DP) has been introduced over a decade ago \cite{Dwork2006, Dwork20061}. Recognized as a robust framework for safeguarding privacy, it's often termed as global DP or centralized DP. DP's strength lies in its mathematical rigor; it operates independent of an adversary's background knowledge and assures potent privacy protection for users. It has found applications across various domains\cite{DP_mec}. However, DP assumes the presence of a trustworthy server, which can be a challenge since many online platforms or crowdsourcing systems might have untrustworthy servers keen on user data statistics\cite{primoff2017equifax, lu2019assessing}.

Emerging from the concept of DP, local differential privacy (LDP) was introduced \cite{Dwork2008}. LDP stands as a decentralized version of DP, offering individualized privacy assurances and making no assumptions about third-party server trustworthiness. LDP has become a focal point in privacy research due to its theoretical significance and practical implications\cite{Rappor}. Numerous corporations, including Apple's iOS\cite{apple_ldp}, Google Chrome, and the Windows operating system, have integrated LDP-driven algorithms into their systems. Owing to its robust capabilities, LDP has become a preferred choice to address individual privacy concerns during various statistical and analytical operations. This includes tasks like frequency and mean value estimation\cite{Tianhao}, the identification of heavy hitters\cite{heavyhitter}, k-way marginal release, empirical risk minimization (ERM), federated learning, and deep learning.

While LDP is powerful, it's not without its challenges, notably in striking an optimal balance between utility and privacy\cite{context}. To address this, there are two primary approaches. Firstly, by devising improved mechanisms - leading to the introduction of numerous LDP-based protocols and sophisticated mechanisms in academic circles. Secondly, by revisiting the definition of LDP itself, with researchers suggesting more flexible privacy concepts to better cater to the utility-privacy balance required for real-world applications. Given the growing significance of LDP, a thorough survey of the topic is both timely and essential. While there exists some literature reviewing LDP, the focus has often been narrow. They either focus on specific applications or certain types of mechanisms.  

In this paper, we delve deep into the world of LDP and its various offshoots, meticulously studying their recent advancements and associated mechanisms. We embark on a thorough exploration of the foundational principles that drive LDP and the evolutionary trajectories of its multiple variants. We aim to identify the cutting-edge developments, shedding light on the innovations that have shaped these privacy tools and the challenges they aim to address in our contemporary digital landscape. Furthermore, we analyze the specific mechanisms that support and enhance the capabilities of LDP, understanding their technical intricacies and the real-world applications they cater to. Through this comprehensive study, we aspire to provide readers with a panoramic view of the current state of LDP research, setting the stage for future inquiries and innovations in this critical domain.

\section{Local Differential Privacy, Properties and Mechanisms}

In this section, we study LDP and its properties, and LDP based mechanisms. We start from the definition of LDP.

\textbf{Definition 1} ($\varepsilon$-Local Differential Privacy ($\varepsilon$-LDP) . \\
A randomized mechanism $M$ satisfies $\varepsilon$-LDP if and only if for any pairs of input values $v$, $v'$ in the domain of $M$, and for any possible output $y \in Y$, it holds
\begin{equation}
P[M(v) = y] \leq e^{\varepsilon} \cdot P[M(v') = y],
\end{equation}
where $P[\cdot]$ denotes probability and $\varepsilon$ is the privacy budget. A smaller $\varepsilon$ means stronger privacy protection, and vice versa.

The basic properties of LDP include the followings:

\textbf{Composition:} \cite{Composition}
Given two mechanisms $M_1$ and $M_2$ that provide $\varepsilon_1$-LDP and $\varepsilon_2$-LDP respectively, their sequential composition provides $(\varepsilon_1 + \varepsilon_2)$-LDP.
\begin{equation}
M(v) = (M_1(v), M_2(v)) \implies M \text{ is } (\varepsilon_1 + \varepsilon_2) \text{-LDP}
\end{equation}

\textbf{Post-processing:} 
Any function applied to the output of an $\varepsilon$-LDP mechanism retains the $\varepsilon$-LDP guarantee.
\begin{equation}
\text{If } M(v) \text{ is } \varepsilon\text{-LDP, then } f(M(v)) \text{ is also } \varepsilon\text{-LDP} .
\end{equation}

\textbf{Robustness to Side Information:} 
LDP guarantees hold even if an adversary has access to auxiliary or side information.

\textbf{Utility-Privacy Tradeoff:} 
Generally, a lower value of $\varepsilon$ implies stronger privacy but might result in reduced utility of the perturbed data.

\textbf{Independence of Background Knowledge:}
The privacy guarantees of LDP mechanisms are designed to hold regardless of any background knowledge an adversary might have.

Next, we study mechanisms that satisfy LDP: 

\textbf{Randomize Response}\cite{random_res}

The Randomized Response Mechanism is a simple yet effective approach to achieving LDP. It's particularly used for binary data, i.e., when a user's data item is either $0$ or $1$. The mechanism operates as follows:

\begin{enumerate}

\item With probability $\frac{1}{2}$, the user truthfully answers a question.
\item With probability $\frac{1}{2}$, the user randomly answers the question.

\end{enumerate}

Mathematically, given a user's true data item $v \in \{0, 1\}$, the mechanism outputs $v$ with probability $\frac{1}{2}$ and outputs $1 - v$ (i.e., the opposite of $v$) with probability $\frac{1}{2}$.

The probability mass function (pmf) is given by:
\begin{align}
P[M(v) = 1] &= \frac{1}{2} v + \frac{1}{2} (1 - v) = \frac{1}{2} \\
P[M(v) = 0] &= \frac{1}{2} (1 - v) + \frac{1}{2} v = \frac{1}{2}
\end{align}

This mechanism ensures $\varepsilon$-LDP with $\varepsilon = \ln(2)$.

\textbf{Laplace Mechanism}\cite{Dwork_DPcontinualOb}

The Laplace Mechanism adds noise drawn from the Laplace distribution to the true value of the data. For LDP, this mechanism can be adjusted as:

Given a data item \( v \), the mechanism outputs:
\[ M(v) = v + \text{Lap}(\frac{\Delta f}{\varepsilon}) \]
where \( \Delta f \) is the sensitivity of the function \( f \) and \( \text{Lap}(\cdot) \) represents the Laplace distribution.

\textbf{Gaussian Mechanism}

Similar to the Laplace Mechanism, the Gaussian Mechanism adds noise but from the Gaussian distribution:

Given a data item \( v \), the mechanism outputs:
\[ M(v) = v + \mathcal{N}(0, \sigma^2) \]
where \( \sigma^2 \) determines the amount of noise based on the desired \( \varepsilon \) and function sensitivity \( \Delta f \).

\textbf{Exponential Mechanism}

The Exponential Mechanism selects an output based on a scoring function and weights outputs with the exponential of their score. Given a set of possible outputs \( R \), a data item \( v \), and a scoring function \( q(v, r) \), the probability of selecting output \( r \) is proportional to:
\[ \exp\left(\frac{\varepsilon q(v, r)}{2\Delta q}\right) \]
where \( \Delta q \) is the sensitivity of \( q \).

\textbf{Perturbed Histogram Mechanism}

For a set of items, instead of perturbing each item, this mechanism perturbs the histogram of the data items. Given a data item set \( V \), the mechanism constructs a histogram \( H \) and then outputs:
\[ M(H) = H + \text{Lap}(\frac{\Delta H}{\varepsilon}) \]
where \( \Delta H \) is the sensitivity of the histogram construction.

Observe that each mechanism's efficacy is closely tied to the sensitivity of the query, denoted as \( \Delta f \). In the realm of Local Differential Privacy (LDP), this sensitivity can often grow significantly, especially when the input domain is vast. The larger the sensitivity, the more noise needs to be introduced by the mechanism to ensure the desired privacy level. This can lead to significant distortion in the data, compromising its utility.

Furthermore, as the input support size increases, maintaining the desired privacy guarantee becomes a challenge. Noise calibrated to a high sensitivity can sometimes overshadow the actual data, rendering the results almost meaningless or leading to misinterpretations.

The consequence of this is a pronounced tradeoff between utility and privacy. Achieving stronger privacy often means accepting reduced accuracy and utility in the results, and vice versa. For applications that require high precision, this can be problematic. It implies that while these mechanisms provide a robust privacy guarantee in theory, their practical applicability can be constrained, especially in scenarios where fine-grained insights from data are crucial. 

Hence, while the promise of LDP is enticing, its real-world implementation requires careful consideration of the utility-privacy balance, pushing researchers to seek more efficient mechanisms or modified privacy models to better cater to practical needs.

\section{Advanced LDP mechanisms }

As we mentioned in the introduction, to improve the utility-privacy tradeoff provided by LDP, there are typically two manners. One is to design dedicated mechanism or advanced protocols. The other is to relax the definition of LDP to enhance the data utility. In this section, we summarized several advanced LDP algorithms, aiming to improve the general utility-privacy tradeoff.

\textbf{RAPPOR \cite{Rappor} (Randomized Aggregatable Privacy-Preserving Ordinal Response):}

Introduced by Google.
 RAPPOR enhances the randomized response mechanism through the incorporation of Bloom filters. Each user's value is hashed multiple times into a Bloom filter, which is then perturbed using the RR technique. This allows multiple string values to be encoded before randomization.
Advantage: Its main strength lies in collecting statistics about low-frequency items in the user population. It can provide meaningful insights even when items are not commonly observed.

\textbf{Local Hashing\cite{Tianhao}:}

Addressing the problem of efficiency in the RR technique when dealing with a large domain of inputs, local hashing maps the original vast domain into a smaller domain using hash functions. This condensed domain can then be analyzed using traditional RR techniques.
Advantage: It substantially reduces the noise introduced in the randomization process, enabling accurate estimation of frequencies for individual items in the domain. This mechanism improves the utility, especially when the original domain is considerably large.

\textbf{Piecewise RR:}

Instead of applying the same randomization mechanism across the entire input domain, the Piecewise RR technique divides the domain into multiple segments or pieces. Each segment then gets its own randomization mechanism tailored to its characteristics.
Advantage: This method achieves a more granular utility-privacy tradeoff. It can offer enhanced privacy in sensitive segments while improving utility in less-sensitive ones.
Optimized RR:

The protocol doesn't just use a fixed randomization parameter; instead, it optimizes the parameters of the RR. This optimization is often based on real data distribution or some auxiliary information, ensuring that the randomization provides the best possible utility.
Advantage: By adjusting the randomization according to data distribution, it achieves better accuracy in aggregate statistics.

\textbf{Fourier Perturbation Algorithm (FPA) \cite{FPA}:}

Instead of perturbing the raw data directly, FPA operates in the frequency domain. The data undergoes a Fourier transformation, after which the perturbation is applied. This allows for randomization in a different space that might be more conducive to certain types of analyses.
Advantage: Provides enhanced utility for specific query types, especially those that are frequency-based or need insights from periodic patterns in data.

\section{LDP variants and mechanisms}
In this section, we introduce LDP variants that aim to provide better utility-privacy tradeoff in different applications.
\subsection{Variants and Mechanisms of LDP}

\subsubsection{\( (\varepsilon, \delta) \)-LDP}
Drawing parallels with how \( (\varepsilon, \delta) \)-DP \cite{DBLP:journals/corr/abs-1810-02810} extends \( \varepsilon \)-DP, \( (\varepsilon, \delta) \)-LDP (sometimes termed as approximate LDP) serves as a more flexible counterpart to \( \varepsilon \)-LDP (or pure LDP). 

\begin{defi}[Approximate Local Differential Privacy]
A randomized process \( M \) complies with \( (\varepsilon, \delta) \)-LDP if, for all input pairs \( v \) and \( v' \) within \( M \)'s domain and any probable output \( y \in Y \), the following holds:
\[ P[M(v) = y] \leq e^{\varepsilon} \cdot P[M(v') = y] + \delta. \]
Here, \( \delta \) is customarily a small value.
\end{defi}

In essence, \( (\varepsilon, \delta) \)-LDP implies that \( M \) achieves \( \varepsilon \)-LDP with a likelihood not less than \( 1-\delta \). If \( \delta = 0 \), \( (\varepsilon, \delta) \)-LDP converges to \( \varepsilon \)-LDP.

\subsubsection{BLENDER Model}
BLENDER \cite{203630}, a fusion of global DP and LDP, optimizes data utility while retaining privacy. It classifies users based on their trust in the aggregator into two categories: the opt-in group and clients. BLENDER enhances utility by balancing data from both. Its privacy measure mirrors that of \( (\varepsilon, \delta) \)-DP \cite{DBLP:conf/eurocrypt/DworkKMMN06}.

\subsubsection{Geo-indistinguishability}
Originally tailored for location privacy with global DP, Geo-indistinguishability \cite{Geo} uses the data's geographical distance. Alvim et al. \cite{DBLP:journals/corr/abs-1805-01456} argued for metric-based LDP's advantages in specific contexts.

\begin{defi}[Geo-indistinguishability]
A randomized function \( M \) adheres to Geo-indistinguishability if, for any input pairs \( v \) and \( v' \) and any output \( y \in Y \), the subsequent relation is met:
\[ P[M(v) = y] \leq e^{\varepsilon \cdot d(v, v')} \cdot P[M(v') = y], \]
where \( d(., .) \) designates a distance metric.
\end{defi}

This model adjusts privacy depending on data distance, augmenting utility for datasets like location or smart meter consumption that are sensitive to distance.

\subsubsection{Local Information Privacy}
Local Information Privacy (LIP) was originally proposed in \cite{Jian1805:Context} as a prior-aware version of LDP, and then, in \cite{jiang2019local}, Jiang et al relax the prior-aware assumption to partial prior-aware (Bounded Prior in their version). The definition of LIP is shown as follows: 

\begin{defi}

$(\epsilon,\delta)$-Local Information Privacy\cite{LIP1}
A mechanism ${M}$  satisfies $(\epsilon,\delta)$-LIP, if $\forall{x\in{\mathcal{X}}}$, $y\in{\textit{Range}(\mathcal{M})}$:
%\begin{equation}\label{def:LIP}
%    e^{-\epsilon}P(Y\in{\mathcal{S}_y})\le{P(Y\in\mathcal{S}_y|X\in\mathcal{S}_x)}\le{e^{\epsilon}}P(Y\in{\mathcal{S}_y})+\delta.
%\end{equation}
\begin{equation}\label{cons0}
\begin{aligned}
    & P(Y=y) \geq e^{-\epsilon}P(Y=y|X=x)-\delta, \\
    &P(Y=y)\le{e^{\epsilon}P(Y=y|X=x)}+\delta.
\end{aligned}
\end{equation}
    
\end{defi}

\subsubsection{Sequential Information Privacy}

Sequential Information Privacy (SIP), built upon LIP, measures the privacy leakage for a data sequence, or time series data. SIP naturally decomposes using chain rule-similar techniques and is comparable to that of LDP.

\begin{defi}

$[(\epsilon)$-Sequence Information Privacy]\cite{jiang2023online}
A mechanism $\mathcal{M}$  satisfies $(\epsilon)$-SIP for some $\epsilon\in{\mathds{R}^+}$, if $\forall{X_1^T\in{\mathcal{X}}}$, $Y_1^T\in{\textit{Range}(\mathcal{M})}$:
\begin{equation}\label{cons0}
\begin{aligned}
    e^{-\epsilon}\le \frac{P[M(x_1^T)=y_1^T]}{P[X_1^T=x_1^T]}\le{e^{\epsilon}}
\end{aligned}
\end{equation}
    
\end{defi}

 \noindent The operational meaning of LIP is, the output $Y$ provides limited additional information about any possible input $X$, and the amount of the additional information is measured by the privacy budget $\epsilon$ and failure probability $\delta$. 

 In \cite{LIP2}, multiple LIP mechanisms were proposed and testified, showing that even though $\epsilon$-LIP is stronger than $2\epsilon$-LDP in terms of privacy protection. The mechanisms achieve more than 2 times of utility gain.
 
\subsubsection{CLDP}
Recognizing LDP's diminished utility with fewer users, Gursoy et al. \cite{DBLP:journals/corr/abs-1905-06361} introduced the metric-based model of condensed local differential privacy (CLDP).

\begin{defi}[\( \alpha \)-CLDP]
For all input pairs \( v \) and \( v' \) in \( M \)'s domain and any potential output \( y \in Y \), a randomized function \( M \) satisfies \( \alpha \)-CLDP if:
\[ P[M(v) = y] \leq e^{\alpha \cdot d(v, v')} \cdot P[M(v') = y], \]
where \( \alpha > 0 \).
\end{defi}

In CLDP, a decline in \( \alpha \) compensates for a growth in distance \( d \). Gursoy et al. employed an Exponential Mechanism variant to devise protocols, particularly benefitting scenarios with limited users.

\subsubsection{PLDP}
PLDP \cite{8368271} offers user-specific privacy levels. Here, users can modify their privacy settings, denoted by \( \varepsilon \).

\begin{defi}[\( \varepsilon \)-PLDP]
For a user \( U \), and all input pairs \( v \) and \( v' \) in \( M \)'s domain and any potential output \( y \in Y \), a randomized function \( M \) meets \( \varepsilon_U \)-PLDP if:
\[ P[M(v) = y] \leq e^{\varepsilon_U} \cdot P[M(v') = y]. \]
\end{defi}

Approaches like the personalized count estimation protocol and advanced combination strategy cater to users with varying privacy inclinations.

\subsubsection{Utility-optimized LDP (ULDP)}

Traditional LDP assumes all data points have uniform sensitivity, often causing excessive noise addition. Recognizing that not all personal data have equivalent sensitivity, the Utility-optimized LDP (ULDP) model was proposed. In this model, let \( KS \subseteq K \) be the sensitive dataset and \( KN = K \setminus KS \) be the non-sensitive dataset. Let \( Y_P \subseteq Y \) be the protected output set and \( Y_I = Y \setminus Y_P \) be the invertible output set. The formal definition of ULDP is:

\begin{defi}
Given \( KS \subseteq K \), \( Y_P \subseteq Y \), a mechanism \( M \) adheres to \( (KS, Y_P, \epsilon) \)-ULDP if:
\begin{itemize}
    \item For every \( y \in Y_I \), there is a \( v \in KN \) with \( P[M(v) = y] > 0 \) and \( P[M(v') = y] = 0 \) for any \( v' \neq v \).
    \item For all \( v, v' \in K \) and \( y \in Y_P \), \( P[M(v) = y] \leq e^{\epsilon} \cdot P[M(v') = y] \).
\end{itemize}
\end{defi}

In simpler terms, \( (KS, Y_P, \epsilon) \)-ULDP ensures that sensitive inputs are mapped only to the protected output set.

\subsubsection{Input-Discriminative LDP (ID-LDP)}

While ULDP classifies data as either sensitive or non-sensitive, the ID-LDP model offers a more nuanced approach by acknowledging varying sensitivity levels among data. It is defined as:

\begin{defi}
Given a set of privacy budgets \( E = \{\epsilon_v\}_{v\in K} \), a mechanism \( M \) adheres to \( E \)-ID-LDP if for all input pairs \( v \) and \( v' \), and any output \( y \in Y \):
\[ P[M(v) = y] \leq e^{r(\epsilon_v, \epsilon_{v'})} \cdot P[M(v') = y] \]
where \( r(\cdot, \cdot) \) is a function of two privacy budgets.
\end{defi}

The study in \cite{DBLP:journals/corr/abs-1807-11317} primarily utilizes the minimum function between \( \epsilon_v \) and \( \epsilon_{v'} \) and introduces the MinID-LDP as a specialized case.

\subsubsection{Parameter Blending Privacy (PBP)}

PBP was proposed as a more flexible LDP variant \cite{10.1145/3320269.3405441}. In PBP, let \( \Theta \) represent the domain of privacy parameters. Given a privacy budget \( \theta \in \Theta \), let \( P(\theta) \) denote the frequency with which \( \theta \) is selected. PBP is defined as:

\begin{defi}
A mechanism \( M \) adheres to \( r \)-PBP if, for all \( \theta \in \Theta, v, v' \in K, y \in Y \), there exists a \( \theta' \in \Theta \) such that:
\[ P(\theta)P[M(v; \theta) = y] \leq e^{r(\theta)} \cdot P(\theta')P[M(v'; \theta') = y] \]
\end{defi}

\subsection{A Summary of LDP variants}
Local Differential Privacy (LDP) is a foundational approach tailored for all data types and operates using the randomized response (RR) technique. Its primary advantage is its broad applicability, but it may add more noise than necessary, especially when not all data attributes have the same sensitivity levels. To address this, approximate LDP, which allows for minor violations in privacy guarantees, introduces flexibility. However, this relaxation can be a double-edged sword, potentially compromising privacy in highly sensitive scenarios.

BLENDER, on the other hand, is crafted explicitly for categorical data. By synergizing aspects of both global Differential Privacy and LDP, it aims to improve data utility. Yet, its reliance on grouping user data might introduce challenges in dynamic or constantly changing environments. Local d-privacy is another variant, designed with metric spaces in mind. It’s particularly beneficial for data like location points, but may not be the first choice for other data structures due to its specific metric-based method.

CLDP stands out for its unique approach to address challenges that arise with smaller user counts, an often overlooked but crucial aspect in privacy. However, while it addresses issues in smaller datasets, it might introduce complexities when the user base grows, making scalability a potential concern. PLDP, meanwhile, strives to provide a more granular level of privacy. While this granularity is its strength, the trade-off might be a more significant computational overhead and intricate implementation details.

ULDP takes a novel stance by focusing on optimizing utility through an emphasis on sensitive data. The premise here is that not all data pieces hold equal sensitivity. However, the challenge and responsibility of correctly categorizing which data is sensitive can be daunting. ID-LDP further refines this concept by providing protection based on the actual sensitivity of the input, using unary encoding to achieve this. Its main challenge is the intricate parameter setting required to ensure optimal performance. Lastly, PBP is distinct in its pursuit of robust privacy. By maintaining the secrecy of provider parameters, it bolsters privacy assurances. Yet, this added layer of secrecy might introduce complexities in implementation and understanding.

\section{Conclusion}

In the realm of data privacy, Local Differential Privacy (LDP) stands out as a vital tool for preserving user data. This research delves into various LDP mechanisms, protocols and variants in definition, each addressing unique challenges. From the foundational LDP to specialized versions like BLENDER for categorical data and Local d-privacy for metrics, the spectrum of solutions is vast. Techniques like CLDP tackle smaller datasets, while PLDP, ULDP, and ID-LDP optimize data utility and privacy levels. The introduction of PBP emphasizes secrecy in privacy parameters. Ultimately, this paper underscores the importance of selecting the right LDP variant, given the specific nature of data and privacy needs.

\bibliographystyle{IEEEtran}
\bibliography{ref,ref2}

\end{document}